\documentclass{svjour2}                    
\smartqed  
\usepackage{graphicx}
\begin{document}

\title{Unconditional no-hidden-variables theorem}



\author{Koji Nagata}


\institute{Department of Physics, Korea Advanced Institute of Science and Technology,
Taejon 305-701, Korea
}

\date{Received: date / Accepted: date}

\maketitle

\begin{abstract}
Recently, [{arXiv:0810.3134}] is accepted and published.
We present ultimate version 
of no-hidden-variables theorem.
We derive a proposition concerning the quantum theory 
under the existence of 
the Bloch sphere in a single spin-1/2 system.
The existence of a single classical probability space for measurement outcome
within the formalism of von Neumann's 
projective measurement does not coexist with the proposition concerning 
the quantum theory.
We have to give up the existence of such a classical 
probability space for measurement outcome in the two-dimensional Hilbert space formalism of the quantum theory.
The quantum 
theory does not accept a hidden-variable interpretation in the two-dimensional
space.
\end{abstract}

\section{Introduction}
Recently, \cite{NagataNakamura} is accepted and published.
As a famous physical theory, the quantum theory
(cf. \cite{JJ,Peres,Redhead,Neumann,NIELSEN_CHUANG}) gives accurate and at times remarkably accurate numerical predictions. Much experimental data 
has fit to the quantum predictions for long time.

On the other hand, from the incompleteness 
argument of Einstein, Podolsky, and Rosen (EPR) \cite{bib:Einstein}, a hidden-variable interpretation
of the quantum theory
has been an attractive topic of 
research \cite{Peres,Redhead}.
There are two main approaches 
to study the hidden-variable interpretation
of the quantum theory.
One is the Bell-EPR theorem \cite{bib:Bell}. 
This theorem says that the quantum predictions violate the inequality following from the EPR-locality condition in the Hilbert space formalism of 
the quantum theory.
The EPR-locality condition tells that a result of measurement pertaining to one system is independent of any measurement performed simultaneously at a distance on another system.

The other is the no-hidden-variables theorem of 
Kochen and Specker (KS theorem) \cite{bib:KS}. 
The original KS theorem says the non-existence of a 
real-valued function which is multiplicative and linear 
on commuting operators.
The quantum theory does not accept the KS type of hidden-variable theory.
The proof of the original KS theorem relies on intricate geometric argument.
Greenberger, Horne, and Zeilinger
discover \cite{bib:GHZ,GHZ} the 
so-called GHZ theorem for four-partite GHZ state.
And, the KS theorem becomes very simple form (see 
also Refs.~\cite{bib:Red,bib:Mermin3,Mermin4,bib:Peres310,bib:Mermin1}).

Mermin considers the Bell-EPR theorem in a multipartite state.
He derives multipartite 
Bell inequality \cite{bib:Mermin2}. 
The quantum predictions by $n$-partite GHZ state violate
the Bell-Mermin inequality by an amount that grows exponentially with $n$.
And, several multipartite Bell inequalities are
reported \cite{Roy,Ardehali,Belinskii,WERNER,Z1,Z2,Z3,W1,W2}.
They also say that the quantum predictions
violate local hidden-variable theories
by an amount that grows exponentially with $n$.

As for the KS theorem, it is begun to research the validity of the KS theorem
by using 
inequalities
(see Refs.~\cite{Simon,Larsson,Cabello,Nagata}).
To find such inequalities to test the validity of the KS theorem 
is particularly useful for 
experimental investigation \cite{bib:experi}.
The KS theorem is 
related to the algebraic structure of a set of quantum operators.
The KS theorem is 
independent of a quantum state under study.
Nagata derives an inequality \cite{Nagata} as tests for the validity of 
the KS theorem.
The quantum predictions violate the Nagata inequality when the system is in
an uncorrelated state.
The uncorrelated state is defined in Ref.~\cite{bib:Werner1}.
The quantum predictions by 
$n$-partite uncorrelated state violate 
the Nagata inequality by an amount that grows exponentially with $n$.

Leggett-type nonlocal hidden-variable theory \cite{Leggett} 
is experimentally investigated  \cite{1,2,3}.
The experiments report that the quantum theory does not accept Leggett-type nonlocal hidden-variable theory. 
These experiments are done in the four-dimensional space (two parties) 
in order to
study nonlocality of hidden-variable theories.
We ask: Can the quantum theory accept a hidden-variable interpretation in the
two-dimensional space (only one party)?
The answer is ``No'' whereas Kochen and Specker explicitly
constructed \cite{bib:KS} such a hidden-variable interpretation in the two-dimensional space formalism of the quantum theory.

Here we aim to show alternative and ultimate version 
of no-hidden-variables theorem.
A pure spin-1/2 state (a quantum state with the two-dimensional space) 
constructs our theorem.
Our theorem says that 
a hidden-variable interpretation of the quantum theory 
from 
the two-dimensional Hilbert space formalism of the quantum theory 
is impossible. 
The quantum theory does not accept a hidden-variable interpretation 
in the two-dimensional space.

In what follows, we derive a proposition concerning the quantum theory under
the existence of the Bloch sphere in a spin-1/2 system. 
The existence of a classical probability space for measurement 
outcome within the formalism of von Neumann's projective 
measurement does not 
coexist with the proposition concerning the quantum theory
under
the existence of the Bloch sphere.
We consider a single classical probability space.
A single classical probability space is enough to 
investigate a hidden-variable interpretation of the quantum theory. 
We can consider 
the direct product of many spaces 
($\Omega_1\times\Omega_2\times\Omega_3\times\cdots$) as a single space.


\section{Notation and preparations }\label{Notation}

Throughout this paper, we assume 
von Neumann's projective measurement.
Throughout this paper,
we confine ourselves to the two-dimensional space.
Let ${\bf R}$ denote the reals where $\pm\infty\not\in {\bf R}$.
We assume that every eigenvalue in this paper lies in ${\bf R}$.
We assume that every Hermitian operator 
is associated with a unique observable (see Ref.~\cite{Redhead}).
We do not need to distinguish between them in this paper.

We investigate if
a hidden-variable interpretation of the two-dimensional Hilbert space formalism of the quantum theory is possible.
Let ${\cal O}$
be the space of Hermitian operators described in the 
two-dimensional Hilbert
space.
Let ${\cal T}$ 
be the space of density operators
 described in the Hilbert
space. Namely, ${\cal T}=
\{\psi | \psi\in{\cal O}\wedge\psi\geq 0\wedge {\rm Tr}[\psi]=1\}$.
We define the notation $\theta$ which represents 
one result of quantum measurements.
Suppose that 
the measurement of 
a Hermitian operator $A$ for a system in the state $\psi$ yields a 
value $\theta(A)\in {\bf R}$.
We consider the following propositions.
We define $\chi_{\Delta}(x), (x\in{\bf R})$ as
the characteristic function.
We define $\Delta$ as any subset of the reals ${\bf R}$.

{\bf Proposition:} BSF ({\it the Born statistical formula}),
\begin{eqnarray}
{\rm Prob}(\Delta)_{\theta(A)}^{\psi}={\rm Tr}[\psi\chi_{\Delta}(A)].
\end{eqnarray}
The symbol $(\Delta)_{\theta(A)}^{\psi}$ denotes the following proposition:
$\theta(A)$ lies in $\Delta$ if the system is in the state $\psi$.
The symbol ``${\rm Prob}$'' denotes the probability that
the proposition $(\Delta)_{\theta(A)}^{\psi}$ holds.

We
consider a classical probability space 
$(\Omega,\Sigma,\mu_{\psi})$.
$\Omega$ is a nonempty space. $\Sigma$ is a
$\sigma$-algebra of subsets of $\Omega$. $\mu_{\psi}$ is a
$\sigma$-additive normalized measure on $\Sigma$ such that 
$\mu_{\psi}(\Omega)=1$.
The subscript $\psi$ means that
the probability measure is determined uniquely
when the quantum state $\psi$ is specified.

We introduce measurable functions 
(classical random variables) onto $\Omega$
($f: \Omega \mapsto {\bf R}$).
The measurable function is written as $f_A(\omega)$
for an operator $A\in {\cal O}$.
Here $\omega\in\Omega$ is a hidden variable.

{\bf Proposition:} HV ({\it a 
hidden-variable interpretation of the quantum theory}).

The measurable function $f_A(\omega)$ exists for 
every Hermitian operator $A$ in ${\cal O}$.

{\bf Proposition:} D ({\it the probability distribution rule}),
\begin{eqnarray}
\mu_{\psi}(f^{-1}_{A}(\Delta))={\rm Prob}(\Delta)_{\theta(A)}^{\psi}.
\end{eqnarray}

The possible value of $f_A(\omega)$ takes eigenvalues of $A$ 
almost everywhere with respect 
to $\mu_{\psi}$ in $\Omega$ if we assign the truth value ``1'' 
for Proposition BSF, Proposition HV, and Proposition D,
simultaneously.
We have the following Lemma.

{\bf Lemma:} 
Let $S_A$ stand for the eigenvalues of the Hermitian operator $A$.
For every quantum state described in a Hilbert 
space, 
\begin{eqnarray}
{\rm BSF}\wedge{\rm HV}\wedge
{\rm D}
\Rightarrow
f_A(\omega)\in S_A,~(\mu_{\psi}-a.e.).\label{MO}
\end{eqnarray}
We review the following:

{\bf Lemma:}\cite{Nagata}  Let $S_A$ stand for the eigenvalues 
of the Hermitian operator $A$.
If
\begin{eqnarray}
{\rm Tr}[\psi A]&:=&\sum_{y\in S_A} 
{\rm Prob}(\{y\})_{\theta(A)}^{\psi}y,\nonumber\\
E_{\psi}(A)&:=&\int_{\omega\in \Omega}\mu_{\psi}({\rm d}\omega)
f_{A}(\omega),\nonumber
\end{eqnarray}
then 
\begin{eqnarray}
{\rm BSF}\wedge{\rm HV}\wedge{\rm D}
\Rightarrow
{\rm Tr}[\psi A]=E_{\psi}(A).\label{QMHV}
\end{eqnarray}

{\it Proof:}
Note
\begin{eqnarray}
&&\omega\in f^{-1}_{A}(\{y\})\Leftrightarrow
f_{A}(\omega)\in \{y\}\Leftrightarrow
y=f_{A}(\omega),\nonumber\\
&&\int_{\omega\in f^{-1}_{A}(\{y\})}
\frac{\mu_{\psi}({\rm d}\omega)}{\mu_{\psi}
(f^{-1}_{A}(\{y\}))}=1,\nonumber\\
&&y\neq y'\Rightarrow 
f^{-1}_{A}(\{y\})\cap f^{-1}_{A}(\{y'\})=\phi.
\end{eqnarray}
Hence we have
\begin{eqnarray}
&&{\rm Tr}[\psi A]=\sum_{y\in S_A}{\rm Prob}(\{y\})_{\theta(A)}^{\psi}y
=\sum_{y\in{\bf R}} 
{\rm Prob}(\{y\})_{\theta(A)}^{\psi}y\nonumber\\
&&=\sum_{y\in{\bf R}}\mu_{\psi}(f^{-1}_{A}(\{y\}))y\nonumber\\
&&=\sum_{y\in{\bf R}}
\mu_{\psi}(f^{-1}_{A}(\{y\}))y
\times \int_{\omega\in f^{-1}_{A}(\{y\})}
\frac{\mu_{\psi}({\rm d}\omega)}{\mu_{\psi}
(f^{-1}_{A}(\{y\}))}\nonumber\\
&&=\sum_{y\in{\bf R}}\int_{\omega\in f^{-1}_{A}(\{y\})}
\mu_{\psi}(f^{-1}_{A}(\{y\}))
\times 
\frac{\mu_{\psi}({\rm d}\omega)}{\mu_{\psi}
(f^{-1}_{A}(\{y\}))}f_{A}(\omega)\nonumber\\
&&=\int_{\omega\in \Omega}\mu_{\psi}({\rm d}\omega)
f_{A}(\omega)=E_{\psi}(A).
\end{eqnarray}
QED.

The probability measure $\mu_{\psi}$ is 
chosen such that the following equation
is valid if we assign the truth value ``1'' for 
Proposition BSF, Proposition HV, and Proposition D, simultaneously:
\begin{eqnarray}
{\rm Tr}[\psi A]=\int_{\omega\in\Omega}\mu_{\psi}({\rm d}\omega)
f_A(\omega)
\end{eqnarray}
for every Hermitian operator $A$ in ${\cal O}$.

\section{Unconditional no-hidden-variables theorem}

We discuss main result of this paper.
Assume a pure spin-$1/2$ state $\psi$ in the two-dimensional space.
Let $\vec \sigma$ be $(\sigma_x,\sigma_y,\sigma_z)$.
$\vec \sigma$ is the vector of Pauli operator.
The measurements (observables) on a pure spin-1/2 state 
of $\vec n\cdot\vec\sigma$ are parameterized by 
a unit vector $\vec n$ in ${\bf R}^{\rm 3}$ (its 
direction along which the spin component is measured).
Here, $\cdot$ is the scalar product in ${\bf R}^{\rm 3}$. 
We define three vectors in ${\bf R}^{\rm 3}$ as 
$\vec x^{(1)}:=\vec x$, $\vec x^{(2)}:=\vec y$, and $\vec x^{(3)}:=\vec z$.
They are the Cartesian axes relative to which spherical angles are measured.
We write the unit vectors in a spherical coordinate system 
 defined by $\vec x^{(1)}$, $\vec x^{(2)}$, and $\vec x^{(3)}$ 
 in the following way:
\begin{eqnarray}
\vec n:=\sin\theta\cos\phi \vec x^{(1)}
+\sin\theta\sin\phi \vec x^{(2)}
+\cos\theta \vec x^{(3)}.
\label{vector}
\end{eqnarray}
We define a quantum expectation value $E_{\rm QM}$ as
\begin{eqnarray}
E_{\rm QM}:= {\rm Tr}[\psi \vec n\cdot \vec \sigma].\label{et}
\end{eqnarray}

We derive a necessary condition for
the quantum expectation value
for the system in a pure spin-1/2 state $\psi$ given in (\ref{et}).
We derive the possible values of the scalar product 
$\int\Omega \left(E_{\rm QM}\times E_{\rm QM}\right)
=:\Vert  E_{\rm QM}\Vert^2$.
$E_{\rm QM}$ is the quantum expectation value given in (\ref{et}).
We use decomposition (\ref{vector}).
We introduce the usual measure 
$\int\Omega=\sin\theta d\theta d\phi$.
We introduce simplified notations as
\begin{eqnarray}
T_{i}=
{\rm Tr}[\psi \vec{x}^{(i)}\cdot \vec \sigma ]
\end{eqnarray}
and
\begin{eqnarray}
(c^1, c^2, c^3)=(\sin \theta\cos\phi,
\sin\theta\sin\phi, \cos\theta).
\end{eqnarray}
We have
\begin{eqnarray}
&&\Vert E_{\rm QM}\Vert^2  = 
\int\Omega
\left(\sum_{i=1}^3T_{i}
c^{i}\right)^2  =  4\pi/3
\sum_{i=1}^3T_{i}^2\leq 4\pi/3,
\label{EEvalue}
\end{eqnarray}
where we use the orthogonality relation
$\int\Omega ~ c_k^{\alpha} c_k^{\beta}  = (4\pi/3) \delta_{\alpha,\beta}$.
$\sum_{i=1}^3T_{i}^2$ is bounded as 
$\sum_{i=1}^3T_{i}^2\leq 1$ if we assign the truth value ``1'' for
a proposition of the quantum theory [The Bloch sphere exists.].
The reason of the condition 
(\ref{EEvalue}) is the Bloch sphere
\begin{eqnarray}
\sum_{i=1}^3 
({\rm Tr}[\psi \vec{x}^{(i)}\cdot \vec \sigma])^2\leq 1.
\end{eqnarray}
We derive a proposition concerning the quantum theory 
while assigning the truth value ``1'' for the quantum proposition 
[The Bloch sphere exists.]
(in a spin-1/2 system). The proposition is $\Vert E_{\rm QM}\Vert^2\le 4\pi/3$.
This inequality is saturated iff $\psi$ is a pure spin-1/2 state.
We derive the following proposition concerning the quantum theory
while assigning the truth value ``1'' for the quantum proposition 
[The Bloch sphere exists.]
\begin{eqnarray}
\Vert E_{\rm QM}\Vert^2_{\rm max}=4\pi/3.\label{Bloch}
\end{eqnarray}
The symbol ``$\Vert E_{\rm QM}\Vert^2_{\rm max}$'' is the maximal value of $\Vert E_{\rm QM}\Vert^2$.

We assign the truth value ``1'' for Proposition BSF, Proposition HV, and Proposition D, simultaneously. 
Then, the quantum expectation value given in (\ref{et}) is
\begin{equation}
E^k_{\rm QM}=\int_{\omega\in\Omega}\mu_{\psi}({\rm d}\omega)
f_{\vec n_k}(\omega).\label{avg}
\end{equation}
The possible values of $f_{\vec n_k}(\omega)$ are $\pm 1$ (in $\hbar/2$ unit) almost everywhere with respect 
to $\mu_{\psi}$ in $\Omega$ if 
we assign the truth value ``1'' for Proposition BSF, Proposition HV, and Proposition D, simultaneously.

We derive a necessary condition for 
the quantum expectation value given in (\ref{avg}).
Again, we derive the possible values of the scalar product 
$\Vert E_{\rm QM}\Vert^2$
of the quantum expectation value. 
The quantum expectation value is $E^k_{\rm QM}$ given in (\ref{avg}).
We have
\begin{eqnarray}
\Vert E_{\rm QM}\Vert^2&=&
\int\Omega
\left(\int_{\omega\in\Omega}\mu_{\psi}({\rm d}\omega)f_{\vec n_k}(\omega)
 \times 
 \int_{\omega'\in\Omega}\mu_{\psi}({\rm d}\omega')f_{\vec n_k}(\omega')
\right)\nonumber\\
&=&\int\Omega
\left(\int_{\omega\in\Omega}\mu_{\psi}({\rm d}\omega)
\int_{\omega'\in\Omega}\mu_{\psi}({\rm d}\omega')
f_{\vec n_k}(\omega)f_{\vec n_k}(\omega')
 \right)\nonumber\\
&\leq &\int\Omega
\left(\int_{\omega\in\Omega}\mu_{\psi}({\rm d}\omega)
\int_{\omega'\in\Omega}\mu_{\psi}({\rm d}\omega')
\left|f_{\vec n_k}(\omega)f_{\vec n_k}(\omega')\right|
 \right)\nonumber\\
&=&\int\Omega
\left(\int_{\omega\in\Omega}\mu_{\psi}({\rm d}\omega)
\int_{\omega'\in\Omega}\mu_{\psi}({\rm d}\omega')
 \right)
=4\pi.\label{inequality2}
\end{eqnarray}
The above inequality (\ref{inequality2}) is saturated since
\begin{eqnarray}
&&\{\omega|\omega\in\Omega\wedge f_{\vec n_k}(\omega)=1,~(\mu_{\psi}-a.e.)\}
\nonumber\\
&&=
\{\omega'|\omega'\in\Omega\wedge f_{\vec n_k}(\omega')=1,~(\mu_{\psi}-a.e.)\},
\nonumber\\
&&\{\omega|\omega\in\Omega\wedge f_{\vec n_k}(\omega)=-1,~(\mu_{\psi}-a.e.)\}
\nonumber\\
&&=
\{\omega'|\omega'\in\Omega\wedge f_{\vec n_k}(\omega')=-1,~(\mu_{\psi}-a.e.)\}.
\end{eqnarray}
Hence we derive the following proposition
if we assign the truth value ``1'' for
Proposition BSF, Proposition HV, and Proposition D, simultaneously
\begin{eqnarray}
\Vert E_{\rm QM}\Vert^2_{\rm max}= 4\pi.\label{BSF}
\end{eqnarray}

We do not assign the truth value ``1'' for two propositions
(\ref{Bloch}) and (\ref{BSF}), simultaneously, 
when the system is in a pure spin-1/2 state.
We are in the contradiction when the system is in a pure spin-1/2 state.

We do not accept the following four propositions, simultaneously, when the system is in a pure spin-1/2 state:
\begin{enumerate}
\item Proposition BSF
\item Proposition HV 
\item Proposition D
\item The Bloch sphere exists.
\end{enumerate}
Suppose that the quantum theory is a set of propositions. 
Suppose that all quantum propositions are true.
We have to give up either Proposition HV or Proposition D
if we assign the truth value ``1'' 
for both Proposition BSF and [The Bloch sphere exists.].
We have to give up a hidden-variable 
interpretation of the two-dimensional 
Hilbert space formalism of the quantum theory.

It is that $\Vert E_{\rm QM}\Vert^2_{\rm max}= 4\pi/3$ if 
we assign the truth value ``1'' for [The Bloch sphere exists.]
when the system is in 
a pure spin-1/2 state.
However, accepting Proposition BSF, Proposition HV, and Proposition D, the existence of a classical probability space of the results of von Neumann's projective measurements assigns 
the truth value ``1'' 
for the different 
proposition $\Vert E_{\rm QM}\Vert^2_{\rm max}=4\pi$.
We are in the contradiction when the system is in a pure spin-1/2 state.
We have to give up, at least, one of propositions,
Proposition BSF,
Proposition HV, 
Proposition D,
[The Bloch sphere exists.], and $\psi$ is a spin-1/2 pure state.

\section{Conclusions}

In conclusion, we have presented alternative and ultimate version 
of no-hidden-variables theorem.
The existence of a classical probability space for the results of von Neumann's projective 
measurement has not coexisted with the existence of the Bloch sphere.  
There has not been a classical probability space for 
projective measurement 
outcome.
Our result has been obtained 
in a quantum system which is in a pure 
spin-1/2 state in the two-dimensional space.
The quantum theory has not accepted a hidden-variable interpretation 
in the two-dimensional space.



\begin{acknowledgements}
This work has been
supported by Frontier Basic Research Programs at KAIST and K.N. is
supported by a BK21 research grant.
\end{acknowledgements}


\begin{thebibliography}{}
%
%

\bibitem{NagataNakamura}
K. Nagata and T. Nakamura,
arXiv:0810.3134.

\bibitem{JJ}
J. J. Sakurai,
{\it Modern Quantum Mechanics}
(Addison-Wesley Publishing Company, 1995),
Revised ed.

\bibitem{Peres}
A. Peres, 
{\it Quantum Theory: Concepts and Methods}
(Kluwer Academic, Dordrecht, The Netherlands, 1993).

\bibitem{Redhead}
M. Redhead,
{\it Incompleteness, Nonlocality, and Realism}
(Clarendon Press, Oxford, 1989), 2nd ed.



\bibitem{Neumann}
J. von Neumann, {\it Mathematical 
Foundations of Quantum Mechanics }
(Princeton University Press, Princeton, New Jersey, 1955).


\bibitem{NIELSEN_CHUANG}
M. A. Nielsen and I. L. Chuang,
\textit{Quantum Computation and Quantum Information}
(Cambridge University Press, 2000).




\bibitem{bib:Einstein}
A. Einstein, B. Podolsky, and N. Rosen, Phys. Rev. {\bf 47}, 777 (1935).






\bibitem{bib:Bell}
J. S. Bell, Physics {\bf 1}, 195 (1964).



\bibitem{bib:KS}
S. Kochen and E. P. Specker,
J. Math. Mech. {\bf 17}, 59 (1967).





\bibitem{bib:GHZ}
D. M. Greenberger, M. A. Horne, and A. Zeilinger,
in {\it Bell's Theorem, Quantum Theory and Conceptions of the Universe},
edited by M. Kafatos (Kluwer Academic, Dordrecht, The Netherlands, 
1989), pp. 69-72.

\bibitem{GHZ}
D. M. Greenberger, M. A. Horne, A. Shimony, and A. Zeilinger,
Am. J. Phys. {\bf 58}, 1131 (1990).




\bibitem{bib:Red}
C. Pagonis, M. L. G. Redhead, and R. K. Clifton,
Phys. Lett. A {\bf 155}, 441 (1991).





\bibitem{bib:Mermin3}
N. D. Mermin,
Phys. Today {\bf 43}(6), 9 (1990).

\bibitem{Mermin4}
N. D. Mermin,
Am. J. Phys. {\bf 58}, 731 (1990).

\bibitem{bib:Peres310}
A. Peres, 
Phys. Lett. A {\bf 151}, 107 (1990).




\bibitem{bib:Mermin1}
N. D. Mermin, Phys. Rev. Lett. {\bf 65}, 3373 (1990).  

\bibitem{bib:Mermin2}
N. D. Mermin, Phys. Rev. Lett. {\bf 65}, 1838 (1990).

\bibitem{Roy}
S. M. Roy and V. Singh,
Phys. Rev. Lett. {\bf 67}, 2761 (1991).
\bibitem{Ardehali}
M. Ardehali, 
Phys. Rev. A {\bf 46}, 5375 (1992).
\bibitem{Belinskii}
A. V. Belinskii and D. N. Klyshko,
Phys. Usp. {\bf 36}, 653 (1993).
\bibitem{WERNER}
R. F. Werner and M. M. Wolf, 
Phys. Rev. A {\bf 61}, 062102 (2000).







\bibitem{Z1}
M. \.Zukowski,
Phys. Lett. A {\bf 177}, 290 (1993).
\bibitem{Z2}
M. \.Zukowski and D. Kaszlikowski, 
Phys. Rev. A {\bf 56}, R1682 (1997).
\bibitem{Z3}
M. \.Zukowski and \v{C}. Brukner, 
Phys. Rev. Lett. {\bf 88}, 210401 (2002).
\bibitem{W1}
R. F. Werner and M. M. Wolf, 
Phys. Rev. A {\bf 64}, 032112 (2001).
\bibitem{W2}
R. F. Werner and M. M. Wolf, 
Quant. Inf. Comp. {\bf 1}, 1 (2001).



\bibitem{Simon}
C. Simon, \v C. Brukner, and A. Zeilinger, 
Phys. Rev. Lett. {\bf 86}, 4427 (2001).
\bibitem{Larsson}
J.-{\AA}. Larsson, Europhys. Lett. {\bf 58}, 799 (2002).
\bibitem{Cabello}
A. Cabello, Phys. Rev. A {\bf 65}, 052101 (2002).

\bibitem{Nagata}
K. Nagata, J. Math. Phys. {\bf 46}, 102101 (2005).


\bibitem{bib:experi}
Y. -F Huang, C. -F. Li, Y. -S. Zhang, J. -W. Pan, and G. -C. Guo,
Phys. Rev. Lett. {\bf 90}, 250401 (2003).


\bibitem{bib:Werner1}
R. F. Werner, 
Phys. Rev. A {\bf 40}, 4277 (1989).

\bibitem{Leggett}
A. J. Leggett,
Found. Phys. {\bf 33}, 1469 (2003).

\bibitem{1}
S. Gr\"oblacher, T. Paterek, R. Kaltenbaek, 
\v{C}. Brukner, M. \.Zukowski, M. Aspelmeyer, and A. Zeilinger,
Nature (London) {\bf 446}, 871 (2007).

\bibitem{2}
T. Paterek, A. Fedrizzi, S. Gr\"oblacher, T. Jennewein, 
M. \.Zukowski, M. Aspelmeyer, and A. Zeilinger,
Phys. Rev. Lett. {\bf 99}, 210406 (2007).

\bibitem{3}
C. Branciard, A. Ling, N. Gisin, C. Kurtsiefer, 
A. Lamas-Linares, and V. Scarani,
Phys. Rev. Lett. {\bf 99}, 210407 (2007).



\end{thebibliography}


\end{document}